\documentclass[pra,twocolumn,amsmath,amssymb,superscriptaddress,unsortedaddress,floatfix]{revtex4}

\usepackage{graphicx}
\usepackage{dcolumn}
\usepackage{bm}
\usepackage{color}
\usepackage{bbm}
\usepackage{soul}

\def\duzomniejsze{<\kern-.7mm<}
\def\duzowieksze{>\kern-.7mm>}

\def\textbf#1{{\bf #1}}
\def\beq{\begin{equation}}
\def\eeq{\end{equation}}
\def\be{\begin{equation}}
\def\ee{\end{equation}}
\def\ben{\begin{eqnarray}}
\def\een{\end{eqnarray}}
\def\beqa{\begin{eqnarray}}
\def\eeqa{\end{eqnarray}}
\def\eea{\end{array}}
\def\bea{\begin{array}}
\newcommand{\bei}{\begin{itemize}}
\newcommand{\eei}{\end{itemize}}
\newcommand{\bee}{\begin{enumerate}}
\newcommand{\eee}{\end{enumerate}}

\begin{document}


\title{Sensitivity of the decay of entanglement of quantum dot
spin qubits to the magnetic field}

\author{Pawe{\l} Mazurek}
\affiliation{National Quantum Information Centre of Gda{\'n}sk, 81-824 Sopot, Poland}
\affiliation{Institute for Theoretical Physics and Astrophysics,
University of Gda{\'n}sk, 80-952 Gda{\'n}sk, Poland}

\author{Katarzyna Roszak}
\affiliation{Institute of Physics, Wroc{\l}aw University of Technology,
50-370 Wroc{\l}aw, Poland}

\author{Ravindra W. Chhajlany}
\affiliation{Faculty of Physics, Adam Mickiewicz University,
61-614 Pozna{\'n}, Poland
}
\affiliation{ICFO - Institut de Ci{\'e}ncies Fot{\'o}niques, Mediterranean Technology Park,
E-08860 Castelldefels, Barcelona, Spain}

\author{Pawe{\l} Horodecki}
\affiliation{National Quantum Information Centre of Gda{\'n}sk, 81-824 Sopot, Poland}
\affiliation{Faculty of Applied Physics and Mathematics, Gda{\'n}sk University of Technology, 
80-952 Gda{\'n}sk, Poland}

\date{\today}

\begin{abstract}
We study the decay of entanglement of quantum dot electron-spin qubits under hyperfine interaction mediated decoherence.
We show that two qubit entanglement of a single entangled initial state may exhibit decay characteristic of the two disentanglement regimes in a single sample, when the external magnetic field is changed. 
The transition is manifested by the supression of time-dependent entanglement
oscillations which are superimposed on the slowly varying entanglement decay related to 
phase decoherence (which result in oscillatory behaviour of entanglement sudden
death time as a function of the magnetic field).
This unique behaviour allows us to propose the double quantum dot two-electron 
spin Bell state as a promising candidate for precise measurements of the magnetic field.
\end{abstract}

\maketitle

Systems of electron spins confined in quantum dots (QDs) have received
much theoretical 
(see Refs \cite{coish2009,coish2010,cywinski2011}
for review) and experimental (see Refs \cite{hanson07a,hanson08,ramsay10}
for review) interest since the initial proposal for spin-based
quantum computing \cite{loss98}. This resulted in the development
of a range of effective techniques for the initialization, manipulation, and 
readout of the spin state in two main trends. One, involving electrical (or magnetic)
manipulation of lateral QDs 
\cite{elzerman04,hanson05,petta05},
and the other, involving optical manipulation of self-assembled QDs 
\cite{kroutvar04,spatzek11,kim11a}.
Both prove sucessful in the generation of high fidelity initial states,
also entangled, but the coherent evolution of spin states and manipulation thereof 
suffer from the destructive effects of the hyperfine interaction 
between the electron spin and 
the spins of the nuclei of the QD atoms. Hence, the current experiments
focus mostly on few-spin qubits
\cite{laird2010,lai2011} which are more robust against decoherence,
or on involved schemes for the minimization of decoherence effects
\cite{greilich07a,barthel12,medford12}.
The proficiency attained in the experiments has been
very recently demonstrated in Ref. \cite{shulman2012}, where quantum
state tomography of two initially entangled singlet-triplet qubits has
been performed.

The central idea of this paper is to propose, based on the high level of 
the experimental techniques used to study electron spin states in QDs, a scheme
for sensing an external parameter (the magnetic field) by harnessing the entanglement present in a two-qubit system and the inbuilt decoherence processes. 
The idea is outside of traditional methods in quantum
metrology, since it relies on decoherence, while metrology requires a high
degree of quantum coherence.
It is vital that the qubits be electron spins confined in QDs 
with a non-zero nuclear spin of the environment, because this 
leads to the specific system-environment interaction and results in a characteristic disentanglement process, which, as we have found, 
strongly and counter-intuitively depends on the magnetic field.

The study of spin entanglement \cite{horodecki2009} decay in a two-electron-two-QD system,
has up-to-date been limited
to a number of complex, yet solvable scenarios \cite{erbe2010,erbe2012b,christ08}.
The complexity accounts for the nontrivial behavior of the reported evolution of entanglement.
Hence, in Refs \cite{erbe2010} and \cite{erbe2012b} the uniform coupling (``box'') model
is extended to account for the exchange interaction between electron spins
for a small number of nuclei in the common
nuclear bath limit (with low bath polarization) and separate nuclear baths limit
(with high bath polarization and large exchange interaction), respectively.
Ref. \cite{christ08} utilizes the ``box'' model with a simplified thermal
spin bath state to introduce a scheme for multipartite entanglement generation
mediated by the interaction with a nuclear bath.
An exception is Ref. \cite{bodoky2009}, where the evolution of entanglement 
of non-interacting spin qubits is studied, but the decoherence model considered is 
phenomenological 
and leads to a different type of decoherence than is reported in the literature 
\cite{cywinski2011,coish2010}.
The importance of Ref. \cite{bodoky2009} lies in its attempt to quantify multipartite entanglement
under a feasibly realistic evolution of the qubit states.

We study the evolution of entanglement of two non-interacting electron spin qubits 
confined in two well separated lateral GaAs QDs. 
The qubits interact via the hyperfine coupling with separate nuclear spin reservoirs, 
which are taken in the high-temperature 
thermalized state to which the baths relax quickly at experimentally accessible
temperatures
\cite{abragam1983}. Hence, we can use the ``box'' model
for the whole range of magnetic field values \cite{merkulov2002, barnes2011},
because entanglement decay takes place on time scales 
shorter than the ``box'' model limit of applicability, $t<N/A$,
where $N$ denotes the number of nuclei, and $A=\sum_k A_k$ is the sum of coupling constants
between the electron and the nuclei.

We show that the nature of entanglement decay changes substantially when the 
transition to the high magnetic field limit is made.
To this end, we study entanglement decay of an initial Bell state, for which sudden death 
of entanglement \cite{zyczkowski01,yu04}
(complete disentanglement while the loss of coherence is still only partial) 
is not possible under pure dephasing processes \cite{roszak06b}.
At high magnetic fields decoherence is restricted to pure dephasing and 
since entanglement is proportional to the coherence,
it decays following the appropriate exponential function.
Contrarily, at low magnetic fields the evolution involves
a redistribution of the spin-up and spin-down occupation.
This leads to entanglement
oscillations which are superimposed on the slowly varying entanglement decay from
phase decoherence, and to entanglement sudden death.
Hence, the same system realizes qualitatively different disentanglement scenarios 
in different magnetic field regimes for the same initial entangled state.

The system can be described by a separable Hamiltonian,
$H=H_{1}\otimes\mathbb{I}_2+\mathbb{I}_1\otimes H_{2}$,
where the individual QD subsystems are described by Hamiltonians of the form 
(the magnetic field is applied in the $z$ direction),
\begin{eqnarray}\label{hamiltonian2}
H_{i}&=&-g\mu_B \hat{S}_i^z B
+\sum_{k} A_{k,i}\hat{S}_i^z \hat{I}_{k,i}^z\\
\nonumber
&&+\frac{1}{2}\sum_{k} A_{k,i}\left(\hat{S}_i^+ \hat{I}_{k,i}^-
+\hat{S}_i^- \hat{I}_{k,i}^+\right),
\end{eqnarray}
with the index $i=1,2$ distinguishing the two dots.
The first term in (\ref{hamiltonian2}) is the electron Zeeman 
splitting, where $g$ is the effective electron g-factor, $\mu_B$ is the Bohr magneton, 
$\hat{S}_i^z$ is the component of the electron spin parallel to the magnetic field, 
and $B$ denotes the applied magnetic field.
The last two terms describe the hyperfine interaction 
between the spin of an electron
and the spins of the surrounding QD nuclei. 
The diagonal (second) term is also known as the Overhauser term and leads to 
pure dephasing, while the last term, known as the ``flip-flop'' term, 
is responsible for both dephasing and leveling out of the electron spin occupations.
Here, $\hat{\mathbf{I}}_{k,i}$ are spin operators
of the individual nuclei (discriminated by the index $k$) in dot $i$. $\hat{I}_{k,i}^z$
is the z-component, while 
$\hat{I}_{k,i}^{\pm}=\hat{I}_{k,i}^{x}\pm i\hat{I}_{k,i}^{y}$ are the nuclear spin raising
and lowering operators. Analogously, $\hat{S}_{i}^{\pm}=\hat{S}_{i}^{x}\pm i\hat{S}_{i}^{y}$
are the raising and lowering operators for the electron spin.
The coupling constants
of the hyperfine interaction depend on the species of the nuclei and on its location
with respect to the electron wave function,
\begin{eqnarray}\label{coupl}
A_{k,i}=A_{k,i}^0 v_0|\Psi_i(\mathbf{r}_{k,i})|^{2},
\end{eqnarray}
where $A_{k,i}^0=\frac{2}{3}\mu_0\gamma_e\gamma_{k,i}$ are the coupling constants
of a given nuclear species found at site $k$ of dot $i$, with
$\mu_0$ denoting the vacuum magnetic permeability, $\gamma_e$ and $\gamma_{k,i}$ being the electron
and nuclear gyro-magnetic ratios, respectively, while $v_0$ is the unit cell volume of the 
QD crystal,
$\Psi_i(\mathbf{r})$ is the wave function of the electron located in dot $i$, 
and $\mathbf{r}_{k,i}$ is the position of the $k$-th 
nucleus in dot $i$.

We have omitted the nuclear Zeeman term and the dipolar interaction between nuclei
in the Hamiltonian (\ref{hamiltonian2}). The first, 
because nuclear Zeeman energies of gallium
and arsenic are very small, and the resulting energy splittings are of the order of tens of 
neV (corresponding to less than a mK) for each Tesla of magnetic field applied to the system. The nearest neighbor dipolar coupling constants between nuclei are even smaller, and
are of the order of $0.1$ neV. Hence, at typical experimental temperatures
both nuclear terms in the Hamiltonian are much smaller than $k_B T$ \cite{merkulov2002,cywinski2011}.

For the same reason the nuclear baths
can be described by infinite-temperature, fully mixed density matrices 
\cite{abragam1983,cywinski2011} unless the state of the nuclear environment is 
especially experimentally prepared.
While a polarized environment
strongly changes the resulting dynamics and leads to an increase of the electron 
spin coherence time \cite{coish2004,zhang2006}, the preparation of such
an environment is demanding experimentally \cite{chekhovich2010,petersen13}
and the currently attainable levels of polarization
are under $70 \%$ \cite{urbaszek2013}. The study of polarized environments is beyond
the scope of this paper, where we wish to describe spin disentanglement 
in the simplest and most common scenario.
Hence, we limit the study to initial states, for which the density matrix of the 
two-qubit subsystem and the two nuclear reservoirs is in the product state
$\varrho(0)=\rho_{DQD}(0)\otimes R_1(0)\otimes R_2(0)$,
where $\rho_{DQD}(0)$ is the initial state of the two confined electron spins,
while the nuclear baths $R_i$ are initially fully mixed.

We use parameters corresponding to two identical lateral GaAs QDs,
but the results are qualitatively valid for any dot type, as long as they can be treated as
non-interacting. Electron wave function envelopes, which are necessary to find the coupling
constants of the hyperfine interaction, Eq.~(\ref{coupl}), are modeled by anisotropic
Gaussians with the extension $l_{\perp}=20$ nm in the $xy$ plane and $l_{z}=2$ nm along $z$ direction,
which is the direction of the applied 
magnetic field.  The number of crystal unit cells considered within each dot  
is $N_1=N_2\approx 1.5\times 10^6$. 

All isotopes naturally found in GaAs
carry spin $I=3/2$ and the nuclear-species-dependent coefficients 
$A_{i,k}^0$
are equal to $A_{Ga^{69}}=36$ $\mu$eV, $A_{Ga^{71}}=46$ $\mu$eV, and
$A_{As^{75}}=43$ $\mu$eV \cite{liu2007,cywinski2009}.
The relative abundances of the gallium isotopes are $60.4 \%$ for $Ga^{69}$ and $39,6\%$
for $Ga^{71}$, this together with the fact that there is one gallium and one arsenic atom
in the GaAs unit cell gives the average hyperfine coupling constant $A=83$ $\mu$eV.
The g-factor is equal to $g=-0.44$ \cite{adachi85}, hence, the
Zeeman electron spin splitting is equal to $25.5$ $\mu$eV
per Tesla of magnetic field.

The parameters are used to find single QD evolutions in the high-magnetic-field limit,
$g\mu_B B \gg A$, for which the ``flip-flop'' term
may be completely neglected.
The condition is fulfilled for magnetic fields
greater than about $3.25$ T.
The Hamiltonian is then diagonal and it is possible to find the evolution for 
a realistic distribution of coupling constants while
taking into account the large number of nuclei.
The resulting dynamics is limited 
to pure dephasing which is further independent of the magnetic field (and
local unitary oscillations that do not disturb entanglement)
for the initial high-temperature environment.
As predicted \cite{merkulov2002}, the decay of a single spin is proportional
to $\exp(-t^2/T^{*2}_2)$, with a characteristic constant 
$T^*_2=\sqrt{\frac{6}{I(I+1)}}\sqrt{N}/A$. 
$\sqrt{N}/A\approx 10$ ns according to the parameters used
and the $T^*_2=12.36$ ns extracted from the calculation
corroborates this.

To quantify single dot evolution at lower magnetic fields, we use the ``box'' model
which is valid on short time scales when the high-temperature nuclear bath density matrix is used
and at high magnetic fields converges with the approach above.
The upper limit of short-time-scale behavior is approximated by $N/A$ \cite{cywinski2009}, 
the value of which is
$1.2\cdot 10^4$ ns for the parameters used and exceeds the disentanglement times by three
orders of magnitude. In the ``box'' model, the hyperfine coupling terms are assumed constant
$A_k=\alpha=A/N$, which allows for exact diagonalization of the Hamiltonian (\ref{hamiltonian2})
as outlined in the Supplementary Materials. Furthermore, ``box'' model evolutions
involving a large number of nuclei can be successfully simulated with reasonably small 
numbers of nuclei, since few-body coherent effects disappear already in the case of $10$ spins-3/2
and for $50$ spins large-number-of-nuclei evolutions are reproduced
(see Supplementary Materials for details). 

The single QD evolution depends strongly on the 
magnetic field. At very low magnetic fields, QD occupations are partially leveled out
due to the interaction with the environment and
phase decoherence closely resembles the decay of the occupation difference. 
The effect of the environment on the occupations
is diminished with growing magnetic field, 
while coherence damping remains strong, although it starts to resemble
exponential decay. When the limit of high magnetic fields
is reached, the interaction with nuclear spins cannot disturb the occupations,
and the pure dephasing process follows a Gaussian decay proportional to
$\exp(-t^2/T^{*2}_2)$.

The study of entanglement evolution 
requires a two-qubit entanglement measure which can be calculated from the system state.
One such measure, for which an explicit formula is available, is the 
concurrence \cite{hill97,wootters98}, which is closely related to the entanglement of 
formation, defined as the ensemble average of the von Neumann 
entropy minimized over all ensemble preparations of the state \cite{bennett96a,bennett96b}.
The concurrence for bipartite entanglement is given by 
$C(\rho_{DQD})=\max\{0,\lambda_{1}-\lambda_{2}-\lambda_{3}-\lambda_{4}\}$, 
where $\lambda_{i}$ are the square roots of the eigenvalues of the matrix 
$\rho_{DQD}(\sigma_{y}\otimes\sigma_{y})\rho_{DQD}^{*}(\sigma_{y}\otimes\sigma_{y})$.
Here, $\rho_{DQD}$ is the two qubit density matrix, $\rho_{DQD}^*$ is its complex conjugate, and
$\sigma_y$ is the appropriate Pauli matrix.

\begin{figure}
\includegraphics[scale=0.7]{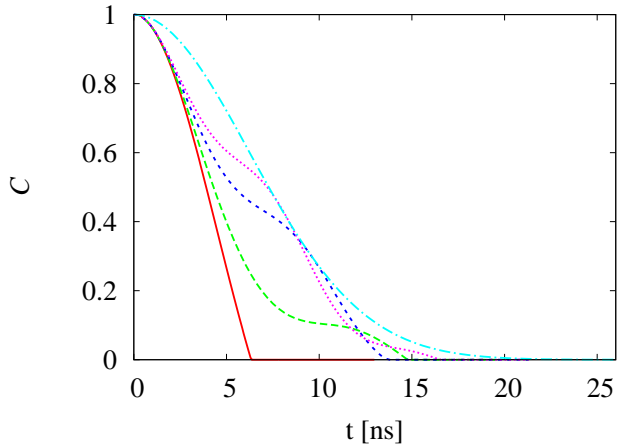}
\caption{\label{ewolucje}
Time evolution of entanglement for different magnetic field values, $B=0$ - solid red line,
$B=11$ mT - long dashed green line, $B=16.5$ mT - dashed blue line, $B=20$ mT - dotted pink line, 
and $B=1$ T - dashed/dotted blue line (high magnetic field limit).}
\end{figure}

We study entanglement evolution of initial maximally entangled
Bell states, 
$|\Psi^{\pm}\rangle=1/\sqrt{2}(|1\rangle \pm|2\rangle)$
and $|\Phi^{\pm}\rangle=1/\sqrt{2}(|0\rangle \pm|3\rangle)$, 
where the states in the single QD basis are equal to
$|0\rangle =|\!\uparrow\uparrow\rangle$, $|1\rangle =|\!\uparrow\downarrow\rangle$,
$|2\rangle =|\!\downarrow\uparrow\rangle$, and $|3\rangle =|\!\downarrow\downarrow\rangle$.
The evolution of the coherences for these initial states
is limited to the single off-diagonal element of 
the density matrix which is initially non-zero, while the other coherences remain zero
at all times. Contrarily, all four occupations are influenced (except for the high-magnetic-field
limit where the decoherence is a pure dephasing process) 
by the interaction.
Hence, the double QD density matrix is simplified and the concurrence
is always given by
\begin{equation}
\label{c}
C(\rho_{DQD})=2\max\{0,
|\rho_{ij}|-\sqrt{\rho_{kk}\rho_{ll}}
\},
\end{equation}
where $i, j$ are equal to $1, 2$ or $0, 3$ depending on the initial state, 
and $k\neq l$, $k\neq i$, $k\neq j$, $l\neq i$, $l\neq j$.
It is evident from Eq.~(\ref{c}) that sudden death of entanglement will occur
when $|\rho_{ij}|<\sqrt{\rho_{kk}\rho_{ll}}$, so it is expected in the low magnetic
field regime when the QD occupations are disturbed, while it will not occur for high
magnetic field pure dephasing.
To the best of our knowledge, such an effect has not been shown neither theoretically 
nor experimentally in any previously studied system. A number of papers showing 
the 
appearance of two regimes of entanglement decay was reported previously, but it was 
either (i) the result of changing the initial state, or (ii) of changing the 
system-environment interaction by changing structural parameters of the qubits 
(which is equivalent to the need of growing a new sample in the laboratory).
Furthermore, because the qubits interact with separate environments at high temperature
thermal equilibrium, the evolution of entanglement is the same for all four
Bell states.

Fig.~\ref{ewolucje} shows entanglement decay
for different magnetic field values.
The zero magnetic field curve (red solid) limits all higher magnetic field curves from
below and ends in sudden death. The high magnetic field curve (dashed/dotted blue line), 
provides the upper limit for the concurrence at a given time
and undergoes exponential decay. In between, the curves corresponding to small magnetic fields
display a more complex entanglement evolution. According to Eq. (\ref{c}), the visible oscillations are due to the interplay of 
the dephasing process and the shifts in the occupations. 
The number of oscillations increases with
the increase of the magnetic field, while they become less pronounced, 
because the high magnetic field inhibits occupation changes. 
The supression of the 
oscillations is a manifestation of the transition between the two types of disentanglement.

\begin{figure}
\includegraphics[scale=0.7]{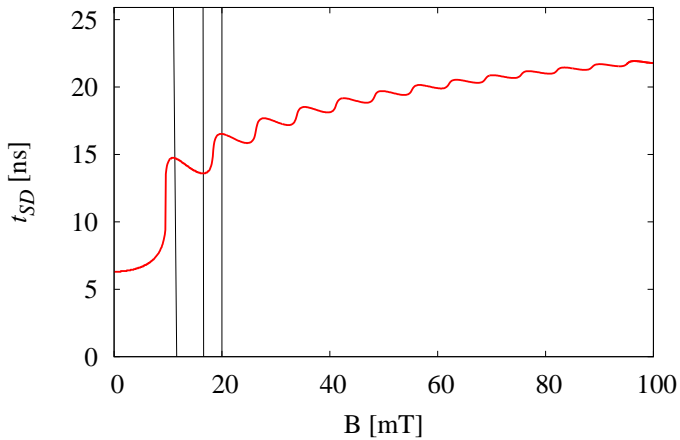}
\caption{\label{death}
Entanglement sudden death time $t_{SD}$ as a function of the magnetic field. The vertical lines mark
the magnetic field values corresponding to the long dashed green,
dashed blue and dotted pink lines in Fig.~\ref{ewolucje}.}
\end{figure}

It is due to those oscillations that the sudden death times 
are not a monotonous function of the magnetic field,
as seen in Fig.~\ref{death}. 
At low magnetic fields, 
a strong oscillatory behavior is evident, starting from around $10$ mT. 
For higher magnetic fields,  $\rho_{ij}$ decays as $\exp[-\sigma^{2}t^{2}]$, 
while $\rho_{kk}=\rho_{ll}$, which initially equals 0, oscillates with the amplitude proportional to $\frac{\sigma^{2}}{B^{2}}$, where $\sigma^{2}\sim A^{2}/N$. 
The equality between the two, responsible for sudden death of entanglement, 
gives an estimated $t_{SD}\sim\sqrt{2 \ln{\frac{B}{\sigma}}}/\sigma$ 
valid for high magnetic fields.

For any fixed time, the considered dynamics is a tensor product of two quantum channels
(completely positive trace preserving maps) which are both 
bi-stochastic - they preserve the maximally mixed state.
The proof of this statement is straightforward, namely:
Consider a single qubit subsystem in a maximally mixed state
coupled by an arbitrary unitary interaction to the
maximally mixed environment state.
Since the product of identities is the identity
operator on the composed (qubit + environment) system
and is invariant under any unitary operation, the
final state of the two subsystems remains the same product
of maximally mixed states. In particular, the maximally mixed state
of the qubit subsystem is preserved.
The argument applies to each of the QDs
separately. It follows that any two-qubit 
state with maximally mixed subsystems will
retain the property during the evolution.
A system with maximally mixed subsystems
has to have Bell diagonal qubit states \cite{horodecki96a}, and
starting from the Bell diagonal state
guarantees Bell diagonality for the whole evolution.
It is known that when a Bell diagonal state becomes separable 
in the course of its evolution, the entanglement fidelity $F$ becomes $1/2$.
We start from a single Bell state (a trivial Bell diagonal state)
and the maximal eigenvalue corresponds 
to the projection onto that state.
Monitoring the difference of that eigenvalue and
$1/2$,  i.e. $W(t)=\frac{1}{2}-F =\frac{1}{2}- \langle\Psi_{0}|\rho(t)|\Psi_{0}\rangle$
(which is a specific entanglement witness), 
we may identify the moment of entanglement sudden death
exactly. This means that the zero point time $t^{*}$ of $W(t)$ 
is just the sudden death time, $t^{*}=t_{SD}$, and
as such has the same dependence on
the magnetic field as shown in Fig. (\ref{death}). 
Quite remarkably, $W$ as an entanglement witness
is directly measurable. In fact, the initialization of a singlet state
(one of the Bell states for spin-up and spin-down qubits)
and the measurement of its Fidelity has been demonstrated in Ref. \cite{petta05}.
By measuring this quantity
we can get the exact estimate of the magnetic field
whenever it corresponds to the initial monotonic regime of
the function. In the regions close to the steep parts of
the function, the above value is quite sensitive to the field $B$ and
can be considered as a threshold sensor of the magnetic field.

We have studied decay of QD spin-qubit Bell state entanglement under decoherence processes mediated
by the hyperfine interaction.
We have shown that varying the magnetic field leads to a transition between
substantially different entanglement decay processes, which is manifested
by the suppression of oscillations in the time-evolution of entanglement. 
Furthermore, at low magnetic fields, the evolution of entanglement displays 
counter-intuitive oscillatory behavior
which results in a non-monotonic dependence of the sudden death time on
the magnetic field. The characteristic behavior is an outcome of the interplay of the decay
of the system coherence with the decoherence induced redistribution of the double QD spin occupations, and can serve as the basis for constructing a threshold magnetic field
sensor utilizing quantum entanglement and quantum decoherence.

\begin{acknowledgments}
The authors thank Bill Coish and {\L}ukasz Cywiński for helpful discussions.
The authors acknowlege support from the
National Science Centre project 2011/01/B/ST2/05459. 
This work was supported by the TEAM programme of the Foundation for Polish Science co-financed from the European Regional Development Fund (K. R.).
P.M. was supported by the Foundation for Polish Science International PhD Projects Programme co-financed by the EU European Regional Development Fund. R.W.C acknowledges funding from the Polish Ministry of Science and Higher Education through a Mobility Plus fellowship.
\end{acknowledgments}


\begin{thebibliography}{10}

\bibitem{coish2009}
W.~A. {Coish} and J. {Baugh}, Phys. Stat. Sol. B {\bf 246},  2203  (2009).

\bibitem{coish2010}
W.~A. {Coish}, J. {Fischer}, and D. {Loss}, \prb {\bf 81},  165315  (2010).

\bibitem{cywinski2011}
{\L}. {Cywi{\'n}ski}, Acta Phys. Pol. A {\bf 119},  576  (2011).

\bibitem{hanson07a}
R. Hanson {\it et~al.}, Rev. Mod. Phys. {\bf 79},  1217–1265  (2007).

\bibitem{hanson08}
R. Hanson and D.~D. Awschalom, Nature {\bf 453},  1043  (2008).

\bibitem{ramsay10}
A.~J. Ramsay, Semicond. Sci. Technol. {\bf 25},  103001  (2010).

\bibitem{loss98}
D. Loss and D.~P. DiVincenzo, Phys. Rev. A {\bf 57},  120  (1998).

\bibitem{elzerman04}
J.~M. Elzerman {\it et~al.}, Nature {\bf 430},  431  (2004).

\bibitem{hanson05}
R. Hanson {\it et~al.}, Phys. Rev. Lett. {\bf 94},  196802  (2005).

\bibitem{petta05}
J.~R. Petta {\it et~al.}, Science {\bf 309},  2180  (2005).

\bibitem{kroutvar04}
M. Kroutvar {\it et~al.}, Nature {\bf 432},  81  (2004).

\bibitem{spatzek11}
S. Spatzek {\it et~al.}, Phys. Rev. Lett. {\bf 107},  137402  (2011).

\bibitem{kim11a}
D. Kim {\it et~al.}, Nature Physics {\bf 7},  223–229  (2011).

\bibitem{laird2010}
E.~A. {Laird} {\it et~al.}, \prb {\bf 82},  075403  (2010).

\bibitem{lai2011}
N.~S. {Lai} {\it et~al.}, Sci. Rep. {\bf 1},  110  (2011).

\bibitem{greilich07a}
A. Greilich {\it et~al.}, Science {\bf 317},  5846  (2007).

\bibitem{barthel12}
C. Barthel {\it et~al.}, Phys. Rev. B {\bf 85},  035306  (2012).

\bibitem{medford12}
J. Medford {\it et~al.}, Phys. Rev. Lett. {\bf 108},  086802  (2012).

\bibitem{shulman2012}
M.~D. {Shulman} {\it et~al.}, Science {\bf 336},  202  (2012).

\bibitem{horodecki2009}
R. {Horodecki}, P. {Horodecki}, M. {Horodecki}, and K. {Horodecki}, Rev. Mod.
  Phys. {\bf 81},  865  (2009).

\bibitem{erbe2010}
B. {Erbe} and J. {Schliemann}, \prb {\bf 81},  235324  (2010).

\bibitem{erbe2012b}
B. {Erbe} and J. {Schliemann}, \prb {\bf 85},  155127  (2012).

\bibitem{christ08}
H. Christ, J.~I. Cirac, and G. Giedke, Phys. Rev. B {\bf 78},  125314  (2008).

\bibitem{bodoky2009}
F. Bodoky, O. G{\"u}hne, and M. Blaauboer, Journal of Physics: Condensed Matter
  {\bf 21},  395602  (2009).

\bibitem{abragam1983}
A. {Abragam}, {\em {The Principles of Nuclear Magnetism}} (Oxford University
  Press, New York, 1983).

\bibitem{merkulov2002}
I.~A. {Merkulov}, A.~L. {Efros}, and M. {Rosen}, \prb {\bf 65},  205309
  (2002).

\bibitem{barnes2011}
E. {Barnes}, {\L}. {Cywi{\'n}ski}, and S. {Das Sarma}, \prb {\bf 84},  155315
  (2011).

\bibitem{zyczkowski01}
K. \ifmmode~\dot{Z}\else \.{Z}\fi{}yczkowski, P. Horodecki, M. Horodecki, and
  R. Horodecki, Phys. Rev. A {\bf 65},  012101  (2001).

\bibitem{yu04}
T. Yu and J.~H. Eberly, Phys. Rev. Lett. {\bf 93},  140404  (2004).

\bibitem{roszak06b}
K. Roszak and P. Machnikowski, Phys. Rev. A {\bf 73},  022313  (2006).

\bibitem{coish2004}
W.~A. {Coish} and D. {Loss}, \prb {\bf 70},  195340  (2004).

\bibitem{zhang2006}
W. Zhang {\it et~al.}, Phys. Rev. B {\bf 74},  205313  (2006).

\bibitem{chekhovich2010}
E.~A. Chekhovich {\it et~al.}, Phys. Rev. Lett. {\bf 104},  066804  (2010).

\bibitem{petersen13}
G. Petersen {\it et~al.}, Phys. Rev. Lett. {\bf 110},  177602  (2013).

\bibitem{urbaszek2013}
B. {Urbaszek} {\it et~al.}, Rev. Mod. Phys. {\bf 85},  79  (2013).

\bibitem{liu2007}
R.-B. Liu, W. Yao, and L.~J. Sham, New Journal of Physics {\bf 9},  226
  (2007).

\bibitem{cywinski2009}
{\L}. {Cywi{\'n}ski}, W.~M. {Witzel}, and S. {Das Sarma}, \prb {\bf 79},
  245314  (2009).

\bibitem{adachi85}
S. Adachi, J. Appl. Phys. {\bf 58},  R1  (1985).

\bibitem{hill97}
S. Hill and W.~K. Wootters, Phys. Rev. Lett. {\bf 78},  5022  (1997).

\bibitem{wootters98}
W.~K. Wootters, Phys. Rev. Lett. {\bf 80},  2245  (1998).

\bibitem{bennett96a}
C.~H. Bennett {\it et~al.}, Phys. Rev. Lett. {\bf 76},  722  (1996).

\bibitem{bennett96b}
C.~H. Bennett, H.~J. Bernstein, S. Popescu, and B. Schumacher, Phys. Rev. A
  {\bf 53},  2046  (1996).

\bibitem{horodecki96a}
R. Horodecki and M. Horodecki, Phys. Rev. A {\bf 54},  1838  (1996).

\end{thebibliography}
\end{document}



\title{Supplementary material: The ``box model''}

\author{Pawe{\l} Mazurek}
\affiliation{National Quantum Information Centre of Gda{\'n}sk, 81-824 Sopot, Poland}
\affiliation{Institute for Theoretical Physics and Astrophysics,
University of Gda{\'n}sk, 80-952 Gda{\'n}sk, Poland}

\author{Katarzyna Roszak}
\affiliation{Institute of Physics, Wroc{\l}aw University of Technology,
50-370 Wroc{\l}aw, Poland}

\author{Ravindra W. Chhajlany}
\affiliation{Faculty of Physics, Adam Mickiewicz University,
61-614 Pozna{\'n}, Poland
}
\affiliation{ICFO - Institut de Ci{\'e}ncies Fot{\'o}niques, Mediterranean Technology Park,
E-08860 Castelldefels, Barcelona, Spain}

\author{Pawe{\l} Horodecki}
\affiliation{National Quantum Information Centre of Gda{\'n}sk, 81-824 Sopot, Poland}
\affiliation{Faculty of Applied Physics and Mathematics, Gda{\'n}sk University of Technology, 
80-952 Gda{\'n}sk, Poland}

\date{\today}

\maketitle

In the ``box model'', the hyperfine coupling terms are assumed constant
$A_k=\alpha=A/N$, which allows for the exact diagonalization of the single dot Hamiltonian 
(Eq.~1 in the main article) as outlined below.

Firstly, it is now possible to rewrite the single dot Hamiltonian 
in terms of the components of the total nuclear spin operator $\hat{K}_{i}=\sum_{k}\hat{I}_{k,i}$.
Furthermore, it is convenient to use the eigenstates of the total nuclear spin and its $z$-component
as the nuclear environment basis states, $\{|K,m\rangle \}$, 
described by the spin quantum numbers $K$ and $m$, $m=-K,-K+1,...,K$, and fulfilling the relations
$\hat{K}_{i}^{2}|Km\rangle=\hbar^{2} K(K+1)|Km\rangle$
and $\hat{K}_{i}^{z}|Km\rangle=\hbar m |Km\rangle$. 
Note that the Hamiltonian (Eq.~1 in the main article) acting on any state $|\sigma;K,m\rangle$,
where $\sigma=\uparrow,\downarrow$ denotes the electron spin, cannot change the nuclear quantum
number $K$ and conserves the $z$-component of the total spin of the combined electron and
nuclear spin system. Hence, 
the Hamiltonian can be represented in easily diagonalizable $2\times 2$ block form, where each block 
links the states
$|\uparrow;K,m\rangle$ and $|\downarrow;K,m+1\rangle$, which form a closed subspace
for every $K$ and $m\in [-K,K-1]$.
The form of these $2\times 2$ blocks is given by
\begin{eqnarray}
\nonumber
\begin{bmatrix}
E_m & M_{K,m}\\
M_{K,m} & -E_{m+1}
\end{bmatrix},
\end{eqnarray} 
where the energies are given by $E_m=\hbar/2(\Omega+\alpha m)$, 
with the Zeeman frequency $\Omega=-g\mu_B B/\hbar$,
and the transitions are governed by $M_{K,m}=\hbar\alpha/2\sqrt{K(K+1)-m(m+1)}$.
The eigenvectors are then of the form
\begin{flalign}
\nonumber
&|+;K,m\rangle=\cos\theta_{K,m}|\uparrow;K,m\rangle+\sin\theta_{K,m}|\downarrow;K,m+1\rangle,\\ 
\nonumber
&|-;K,m\rangle=-\sin\theta_{K,m}|\uparrow;K,m\rangle+\cos\theta_{K,m}|\downarrow;K,m+1\rangle,
\end{flalign}
with 
\begin{equation}
\label{theta}
\sin\theta_{K,m}=\frac{M_{K,m}}{(E^{+}_{K,m}+E_{m+1})^{2}+M_{K,m}^{2}},
\end{equation}
and with corresponding eigenvalues given by
\begin{equation}
\label{eigene}
E^{\pm}_{K,m}=\frac{-\hbar\alpha/2\pm\sqrt{(E_m+E_{m+1})^{2}+4M_{K,m}^{2}}}{2}.
\end{equation}
Furthermore, the $|\!\uparrow;K,K\rangle$ and $|\!\downarrow; K, -K\rangle$ states are eigenvectors
of the Hamiltonian.
Hence, the evolution of any combined state of the electron and nuclear spins can be found.

The high-temperature nuclear spin density matrix, rewritten in the $\{|K,m\rangle \}$ basis
takes the diagonal form
\begin{equation}\label{R}
R(0)=\sum_{K,m}P_{K,m}|K,m\rangle\langle K,m|,
\end{equation}
where $P_{K,m}$ are coefficients describing the multiplicity of the occupation of each state
and satisfy the relation $\sum_{K,m}P_{K,m}=1$.
For nuclei with spin $s$ they are given by \cite{mikhailov1977}
\begin{equation}
P_{K,m}\sim\sum_{i}(-1)^{i}{N\choose i}{(s+1)N-(2s+1)i-K-2\choose N-2},
\end{equation}
where $i\in [0,N]$ is an integer. 
For spin $\frac{1}{2}$ systems the formula simplifies to 
\begin{eqnarray}\label{spiny12}
P_{K,m}\sim\frac{N!(2K+1)}{(\frac{1}{2}N-K)!(\frac{1}{2}N+K+1)!}.
\end{eqnarray}
  
\begin{figure}
\includegraphics[scale=0.66]{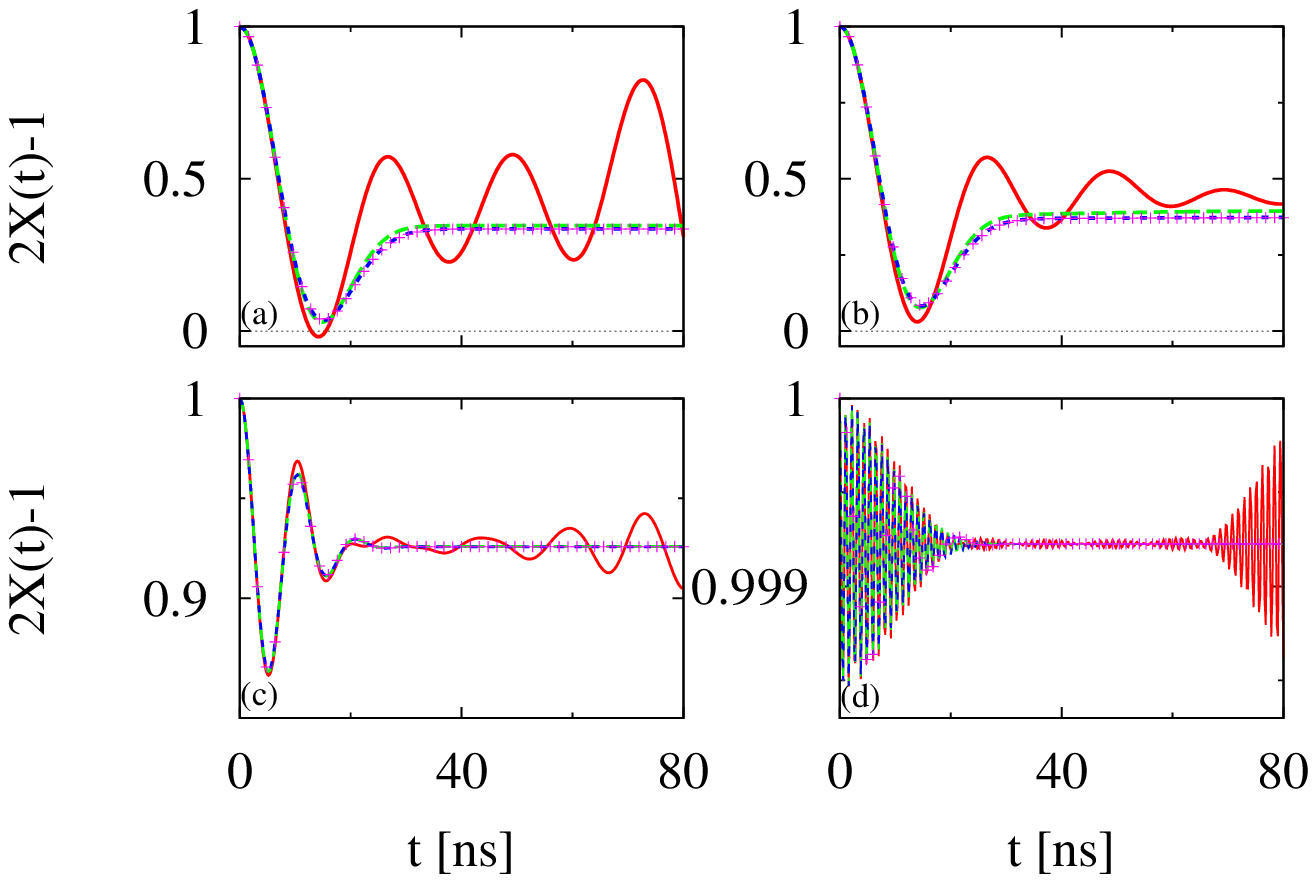}
\caption{\label{obsadzenia} 
``Box model'' evolution of the difference of single QD occupations $\Delta\rho(t)/\Delta\rho(0)$ (see text)
for $B =0$ mT (a), $1.5$ mT (b), $15$ mT (c), and $150$ mT (d). Different curves correspond to
different numbers of spin-3/2 environment nuclei, 
$N=2$ (red, solid line), $10$ (green,dashed), and $50$ (blue, dotted). Points denote the
evolution fitted with $300$ nuclei with spin 1/2.
}
\end{figure}

The evolution of the QD density matrix can be found by tracing out the environmental degrees of 
freedom and assuming a product initial state 
$\rho(0)=\rho_{QD}(0)\otimes R(0)$, with the initial QD density matrix denoted by
$\rho_{QD}(0)$ and the nuclear spin density matrix $R(0)$ 
given by eq. (\ref{R}). It is described by
\begin{eqnarray}
\nonumber
\rho_{QD}(t)&=&|\uparrow\rangle\langle\uparrow|\left(\rho_{\uparrow\uparrow}(0)X(t)
+\rho_{\downarrow\downarrow}(0)(1-X'(t))\right)\\
\nonumber
&&+|\downarrow\rangle\langle\downarrow|\left(\rho_{\uparrow\uparrow}(0)(1-X(t))
+\rho_{\downarrow\downarrow}(0)X'(t)\right)\\
\label{rozwiazanie}
&&+|\uparrow\rangle\langle\downarrow|\left(\rho_{\uparrow\downarrow}(0)Y(t)\right)
+H.c..
\end{eqnarray}
Here, $X(t)=\sum_{K,m}P_{K,m}|X_{K,m}(t)|^2$, $X'(t)=\sum_{K,m}P_{K,m}|X_{K,m-1}(t)|^2$,
and $Y(t)=\sum_{K,m}P_{K,m}X_{K,m}(t)X_{K,m-1}(t)$, with
\begin{equation}
\nonumber
X_{K,m}(t)=\cos^2\theta_{K,m}e^{-\frac{i}{\hbar}\phi_{K,m}t}
+\sin^2\theta_{K,m}e^{\frac{i}{\hbar}\phi_{K,m}t}.
\end{equation}
The phase coefficients are equal to
\begin{equation}
\nonumber
\phi_{K,m}=\frac{\hbar\sqrt{\Omega^2+\Omega\alpha(2m+1)+\frac{\alpha^2}{4}+\alpha^2K(K+1)}}{2},
\end{equation}
and the mixing angles $\theta_{K,m}$ are given by eq. (\ref{theta}).
Note that for a large number of nuclei $N$ the time-dependent functions $X(t)$ and $X'(t)$
coincide, and the evolution of the diagonal terms of the density matrix is described by
\begin{equation}
\label{deltaro}
\Delta\rho(t)=\Delta\rho(0)(2X(t)-1),
\end{equation}
with $\Delta\rho(t)=\rho_{\uparrow\uparrow}(t)-\rho_{\downarrow\downarrow}(t)$.

\begin{figure}
\includegraphics[scale=0.66]{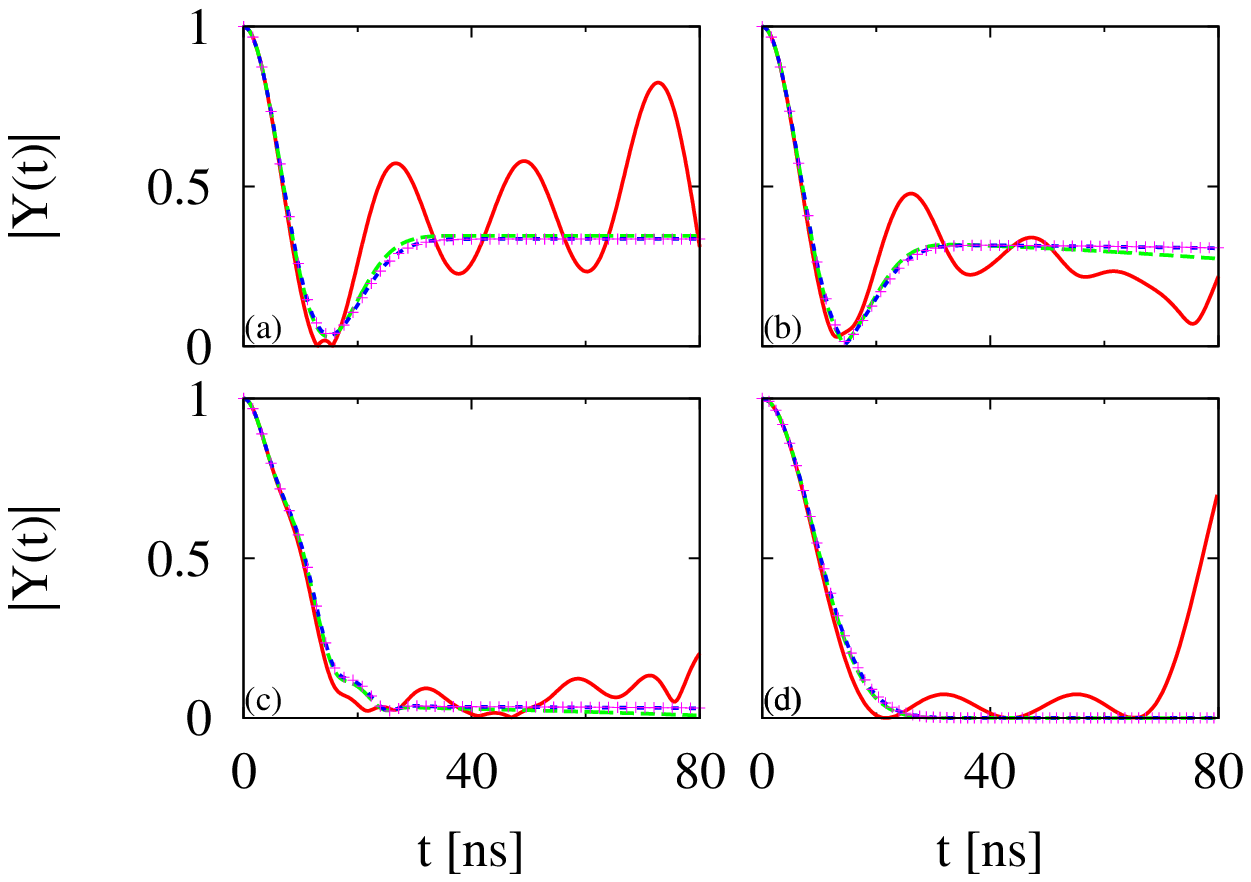}
\caption{\label{koherencje}
``Box model'' evolution of the amplitude of 
single QD coherence $\rho_{\uparrow\downarrow}(t)/\rho_{\uparrow\downarrow}(0)$ (see text)
for $B =0$ mT (a), $1.5$ mT (b), $15$ mT (c), and $150$ mT (d). Different curves correspond to
different numbers of spin-3/2 environment nuclei, 
$N=2$ (red, solid line), $10$ (green,dashed), and $50$ (blue, dotted). Points denote the
evolution fitted with $300$ nuclei with spin 1/2.}
\end{figure}

Even though the ``box model'' is exactly solvable as reproduced above, finding the actual QD
evolution when the environment is in the high-temperature equilibrium state becomes numerically
challenging very quickly with the growing number of nuclei $N$, due to the involved summation
over $K$ and $m$, and is practically impossible for a realistically large values of $N$.
The difficulty of the task also grows rapidly with higher spins of the nuclear 
species taken into account.
Figs.~\ref{obsadzenia} and \ref{koherencje} serve to demonstrate that ``box model'' evolutions
involving a large number of nuclei can be successfully simulated with reasonably small 
numbers of nuclei, since the few-body coherent effects disappear already in the case of $10$ spins-3/2
and for $50$ spins the large number of nuclei evolutions are reproduced.
The necessary condition to achieve convergence is that $A/\sqrt{N}$ remains constant. This requirement stems from the semi-classical approximation \cite{merkulov2002} result giving the characteristic
decay time $T^*_2=\sqrt{\frac{6}{I(I+1)}}\sqrt{N}/A$, where I is the nuclear spin.  

Fig.~\ref{obsadzenia} contains plots of the function $1-2X(t)$ which determines the evolution
of the QD occupations for large $N$ (see eq.~(\ref{deltaro})) for four different magnetic field
values as a function of time. We have found that the function $X'(t)$ converges 
for similar values of $N$ as the function $X(t)$, hence no additional plots are necessary.
Analogously, the plots of the evolution of the amplitude of the off-diagonal terms, 
$|Y(t)|=|\rho_{\uparrow\downarrow}(t)/\rho_{\uparrow\downarrow}(0)|$ are shown
in Fig.~\ref{koherencje}
for the same values of the magnetic field.
In each plot, there are three curves corresponding to spin-3/2 nuclei, with 
$N=2$, $10$ and $50$.
We have also found the evolutions for $N=500$ and $N=2000$,
but the resulting curves are indistinguishable from the $50$ nuclei curves
on relevant time scales and, hence,
have not been included in the plots.
As can be seen, 
the effects resulting from a limited number of spins are strong only for very small $N$.
They manifest themselves as additional oscillations of the 2 nuclei occupation curves
and echo-like characteristics resulting from the alignment of the few nuclear spins
at certain time intervals.

The type of evolution manifested by the single QD system depends strongly on the value of the 
magnetic field. At very low magnetic fields, the occupations of the QD are redistributed 
due to the interaction with the environment, see Fig.~\ref{obsadzenia} (a) and (b).
In this regime, the phase decoherence closely resembles the decay of the occupation difference, 
Fig.~\ref{koherencje} (a), (b). The effect of the environment on the occupations
is diminished with growing magnetic field, Fig.~\ref{obsadzenia} (c), (d), 
while coherence damping remains strong, although it starts to resemble
exponential decay, Fig.~\ref{koherencje} (c), (d). When the limit of high magnetic fields
is reached, the interaction with nuclear spins cannot disturb the QD occupations,
and the pure dephasing process follows a Gaussian decay proportional to
$\exp(-t^2/T^{*2}_2)$.

Analysis of spin-1/2 environments further simplifies the generation of QD system
evolutions for a given number of nuclei. 
To quantify the applicability of such an approximation, points have
been added to the plots of Figs.~\ref{obsadzenia} and \ref{koherencje}, which
denote the evolutions found by modeling the spin 3/2 environment with a spin 1/2 
nuclei for $N=300$. The large number of nuclei is necessary to achieve convergence for spins 1/2.
The fitting required a scaling of the constants $\alpha$ by
$\sqrt{\frac{I_{3/2}(I_{3/2}+1)}{I_{1/2}(I_{1/2}+1)}}=\sqrt{5}$, with $I_{3/2}=\frac{3}{2}$, $I_{1/2}=\frac{1}{2}$, deduced from semi-classical approximation. The resulting evolutions are qualitatively and quantitatively reproduced very well 
and the transitions
between different types of decoherence with growing magnetic field are the same
as in the case of spin-3/2 environment. Unfortunately, the large number of environment
atoms required diminishes any computational advantages, which would be gained
by using nuclei with smaller spins.